\documentclass[fleqn,twoside]{article}
\usepackage{espcrc2}
\usepackage{graphicx}
\usepackage[figuresright]{rotating}
\usepackage{psfig}

\title{Scientific Highlights of the HETE-2 Mission}

\author{
D. Q. Lamb\address[uofc]{Department of Astronomy \& Astrophysics, 
	University of Chicago, Chicago, IL 60637, USA}, 
G. R. Ricker\address[mit]{MIT Center for Space Research, 
	Cambridge, MA 02139, USA}, 
J-L. Atteia\address[cesr]{Centre D'Etude Spatiale des Rayonnements, France}, 
C. Barraud\addressmark[cesr], 
M. Boer\addressmark[cesr], 
J. Braga\address[inpe]{Instituto Nacional de Pesquisas Espaciais,
Sao Jose dos Campos, 12227-010, Brazil}, 
N. Butler\addressmark[mit], 
T. Cline\address[goddard]{NASA Goddard Space Flight Center, Greenbelt,
	MD 20771}, 
G. B. Crew\addressmark[mit], 
J.-P. Dezalay\addressmark[cesr], 
T. Q. Donaghy\addressmark[uofc], 
J. P. Doty\addressmark[mit], 
A. Dullighan\addressmark[mit], 
E. E. Fenimore\address[lanl]{Los Alamos National Laboratory, 
	Los Alamos, NM, USA}, 
M. Galassi\addressmark[lanl], 
C. Graziani\addressmark[uofc], 
K. Hurley\address[ucb]{UC Berkeley, Space Sciences Laboratory, Berkeley, CA 94720, USA}, 
J. G. Jernigan\addressmark[ucb], 
N. Kawai\address[titech]{Tokyo Institute of Technology, Tokyo, Japan}, 
A. Levine\addressmark[mit], 
R. Manchanda\address[tata]{Department of Astronomy and Astrophysics,
	Tata Institute, 
	Mumbai, 400 005, India}, 
M. Matsuoka\address[nasda]{NASDA, Tokyo, Japan}, 
F. Martel\addressmark[mit], 
G. Monnelly\addressmark[mit], 
G. Morgan\addressmark[mit], 
J.-F. Olive\addressmark[cesr], 
G. Pizzichini\address[cndr]{Consiglio Nazionale Delle Ricerche, Italy}, 
G. Prigozhin\addressmark[mit], 
T. Sakamoto\addressmark[titech], 
Y. Shirasaki\address[naoj]{National Astronomical Observatory, Tokyo, Japan}, 
M. Suzuki\addressmark[titech], 
K. Takagishi\address[miyazaki]{Faculty of Engineering, Miyazaki
	University, Gakuen Kibanadai Nishi, Miyazaki, 889-2192, Japan}, 
T. Tamagawa\address[riken]{Institute of Physical and Chemical Research (RIKEN), 
	Tokyo, Japan}, 
K. Torii\addressmark[riken], 
R. Vanderspek\addressmark[mit], 
G. Vedrenne\addressmark[cesr], 
J. Villasenor\addressmark[mit], 
S. E. Woosley\address[ucsb]{Department of Astronomy \& Astrophysics, 
	University of California, Santa Cruz, CA 95064, USA},  
M. Yamauchi\addressmark[miyazaki] and
A. Yoshida\address[aoyama]{Aoyama University, Tokyo, Japan}
}
       
\begin{document}

\begin{abstract}
The HETE-2 mission has been highly productive.  It has observed more 
than 250 GRBs so far.  It is currently localizing 25 - 30 GRBs per
year, and has localized 43 GRBs to date.  Twenty-one of these
localizations have led to the detection of X-ray, optical, or radio
afterglows, and as of now, 11 of the bursts with afterglows have known
redshifts.  HETE-2 has confirmed the connection between GRBs and Type
Ic supernovae, a singular achievement and certainly one of the
scientific highlights of the mission so far.  It has provided evidence
that the isotropic-equivalent energies and luminosities of GRBs are 
correlated with redshift, implying that GRBs and their progenitors
evolve strongly with redshift.  Both of these results have profound
implications for the nature of GRB progenitors and for the use of GRBs
as a probe of cosmology and the early universe.  HETE-2 has placed
severe constraints on any X-ray or optical afterglow of a short GRB. 
It is also solving the mystery of ``optically dark'' GRBs, and
revealing the nature of X-ray flashes.
\vspace{1pc}
\end{abstract}

\maketitle

\section{Introduction}

Gamma-ray bursts (GRBs) are the most brilliant events in the Universe.
Long regarded as an exotic enigma, they have taken center stage in
high-energy astrophysics by virtue of the spectacular discoveries of
the past six years.  It is now clear that they also have important
applications in many other areas of astronomy: GRBs mark the moment of
``first light'' in the universe; they are tracers of the star
formation, re-ionization, and metallicity histories of the universe;
and they are laboratories for studying core-collapse supernovae.

Three major milestones have marked this journey.  In 1992, results from
the Burst and Transient Source Experiment (BATSE) on board the {\it
Compton Gamma-Ray Observatory} ruled out the previous paradigm (in
which GRBs were thought to come from a thick disk of neutron stars in
our own galaxy, the Milky Way), and hinted that the bursts might be
cosmological \cite{meegan1993}.  In 1997, results made possible by {\it
Beppo}SAX \cite{costa1997} decisively determined the distance scale to
long GRBs (showing that they lie at cosmological distances), and
provided circumstantial evidence that long bursts are associated with
the deaths of massive stars (see, e.g., \cite{lamb2000}).  In 2003,
results made possible by the High Energy Explorer Satellite 2 (HETE-2)
\cite{vanderspek2003a} dramatically confirmed the GRB -- SN connection
and firmly established that long bursts are associated with Type Ic
core collapse supernovae.  Thus we now know that the progenitors of
long GRBs are massive stars.

The HETE-2 mission has been highly productive in addition
to achieving this breakthrough:

\begin{itemize}

\item 
HETE-2 is currently localizing 25 - 30 GRBs per year;

\item 
HETE-2 has accurately and rapidly localized 43 GRBs in 2 1/2 years of
operation (compared to 52 GRBs localized by {\it Beppo}SAX during its
6-year mission); 14 of these have been localized to $< 2$ arcmin
accuracy by the SXC plus WXM.

\item 
21 of these localizations have led to the identification of the X-ray,
optical, or radio afterglow of the burst.

\item  
As of the present time, redshift determinations have been reported for
11 of the  bursts with afterglows (compared to 13 {\it Beppo}SAX bursts
with redshift determinations).

\item 
HETE-2 has detected 16 XRFs so far (compared to 17 by {\it Beppo}SAX).

\item
HETE-2 has observed 25 bursts from the soft gamma-ray repeaters
1806-20 and 1900+14 in the summer of 2001; 2 in the summer of 2002; and
18 so far in 2003.  It has discovered a possible new SGR: 1808-20.

\item
HETE-2 has observed $\sim$ 170 X-ray bursts (XRBs) in the summer of
2001, $>$ 500 in the summer of 2002, and $>$ 150 so far in 2003 from
$\sim$ 20 sources.  (We pointed HETE-2 toward the Galactic plane during
the summer of 2002 and caught a large number of XRBs in order to 
calibrate new SXC flight software.)

\end{itemize}

Fourteen GRBs have been localized by the HETE-2 WXM plus SXC so far. 
Remarkably, all 14 have led to the identification of an X-ray, optical,
infrared, or radio afterglow; and 13 of 14 have led to the
identification of an optical afterglow.  In contrast, only $\approx$
35\% of {Beppo}SAX localizations led to the identification of an
optical afterglow.

% Rebrightenings of the GRB 030329 optical afterglow
\begin{figure}[t]
\centerline{\psfig{file=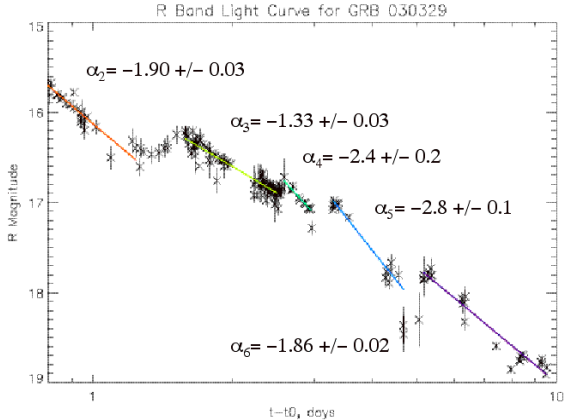,width=0.47 \textwidth}} 
\caption{Successive rebrightenings of the optical afterglow of GRB
030329 during the 10 days following the burst.  From
\cite{filippenko2003}.
\label{fig3}}
\end{figure}

% Discovery spectrum of SSN 2003dh.
\begin{figure}[t]
\centerline{\psfig{file=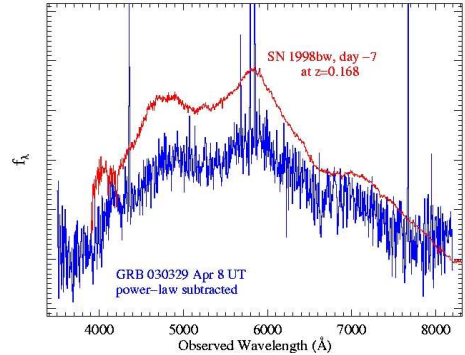,width=0.47 \textwidth}} 
\caption{Comparison of the discovery spectrum of SN 2003dh seen in the 
afterglow of GRB 030329 at 8 days after the burst and the spectrum of
the Type Ic supernova SN 1998bw.  The similarity is striking.  From
\cite{stanek2003}.
\label{fig4}}
\end{figure}

\section{Scientific Highlights of the HETE-2 Mission}

Confirmation of the GRB -- SN connection is a singular achievement and
certainly one of the scientific highlights of the HETE-2 mission. 
Other highlights of the mission include the following:

\begin{itemize}

\item 
HETE-2 made possible rapid follow-up observations of a short GRB,
allowing severe constraints to be placed on the brightness of any
X-ray or optical afterglow.  

\item
The rapid follow-up observations made possible by HETE-2 have opened
the era of high-resolution spectroscopy of optical afterglows (e.g,
GRBs 020813, 021004, and 030329).

\item
Accurate, rapid HETE-2 localizations sent to ground-based robotic
telescopes have made it possible to explore the previously unknown
behavior of optical afterglows in the 3 - 20 hour ``gap'' immediately
following the burst that existed in the {\it Beppo}SAX era.  This has
confirmed the existence of a very bright, distinct phase lasting
$\approx$ 10 minutes.
%, which has been interpreted as emission from a reverse shock 
%produced when the GRB jet slams into circumburst matter.

\item 
HETE-2 is solving the mystery of ``optically dark'' GRBs.  As already
remarked upon, the identification of an optical afterglow for 13 of 14
GRBs localized by the SXC plus WXM instruments on HETE-2 has shown that
very few long GRBs are truly "optically dark."  

\item 
Optical and NIR follow-up observations made possible by HETE-2 have
provided the best case to date of a GRB whose optical afterglow has
been extinguished by dust, and several examples of GRBs with
exceptionally dim optical afterglows.  These GRBs would very likely
have been classified as ``optically dark'' were it not for the
accurate, rapid localizations provided by HETE-2.  

\item 
HETE-2 is revealing the nature of X-ray flashes (XRFs).  Specifically,
HETE-2 has provided strong evidence that the properties of XRFs,
X-ray-rich GRBs, and GRBs form a continuum, and therefore that these
three types of bursts are the same phenomenon.  

\item
HETE-2 results also show that XRFs may provide unique insights into the
nature of GRB jets, the rate of GRBs, and the role of GRBs in Type Ic
supernovae.  In particular, the HETE-2 results provide evidence that
GRB jets are uniform rather than structured.  They also suggest that
the jets are very narrow, and that the rate of GRBs may be much larger
than has been thought.

\end{itemize}

\section{GRB -- SN Connection}

There has been increasing circumstantial and tantalizing direct
evidence in the last few years that GRBs are associated with core
collapse supernovae (see, e.g. \cite{lamb2000}).  The detection and
localization of GRB 030329 by HETE-2 \cite{vanderspek2003a} led to a
dramatic confirmation of the GRB -- SN connection.  GRB 030329 was
among the brightest 1\% of GRBs ever seen.  Its  optical
afterglow was $\sim 12^{\rm th}$ magnitude at 1.5 hours after the burst
\cite{price2003} -- more than 3 magnitudes brighter than the famous
optical afterglow of GRB 990123 at a similar time \cite{akerlof1999}. 
In addition, the burst source and its host galaxy lie very nearby, at a
redshift $z = 0.167$ \cite{greiner2003}.  Given that GRBs typically
occur at $z$ = 1-2, the probability that the source of an observed
burst should be as close as GRB 030329 is one in several thousand.  It
is therefore very unlikely that HETE-2, or even {\it Swift}, will see
another such event.

The fact that GRB 030329 was very bright spurred the astronomical
community -- both amateurs and professionals -- to make an
unprecedented number of observations of the optical afterglow of this
event.  Figure 1 shows the light curve of the optical afterglow of GRB
030329  1-10 days after the burst.  At least four dramatic
``re-brightenings'' of the afterglow are evident in the saw-toothed
lightcurve.  These may be due to repeated injections of energy into the
GRB jet by the central engine at late times, or caused by the
ultra-relativistic jet ramming into dense blobs or shells of material
\cite{granot2003}.  If the former, it implies that the central engine
continued to pour out energy long after the GRB was over; if the
latter, it likely provides information about the last weeks and days of
the progenitor star.

The fact that GRB 030329 was very nearby made its optical afterglow an
ideal target for attempts to confirm the conjectured association
between GRBs and core collapse SNe.  Astronomers were not disappointed:
about ten days after the burst, the spectral signature of an energetic
Type Ic supernova emerged \cite{stanek2003}.  The supernova has been
designated SN 2003dh.  Figure 2 compares the discovery spectrum of SN
2003dh in the afterglow light curve of GRB 030329 and the spectrum of
the Type Ic supernova SN 1998bw.  The similarity is striking.  The
breadth and the shallowness of the absorption lines in the spectra of
SN 2003dh imply expansion velocities of $\approx$ 36,000 km s$^{-1}$ --
far higher than those seen in typical Type Ic supernovae, and higher
even than those seen in SN 1998bw.  It had been conjectured that GRB
980425 was associated with SN 1998bw (see, e.g., \cite{galama1998}),
but the fact that, if the  association were true, the burst would have
had to have been $\sim 10^4$ times fainter than any other GRB observed
to date made the association suspect.  The clear detection of SN 2003dh
in the afterglow of GRB 030329 confirmed decisively the connection
between GRBs and core collapse SNe.

The association between GRB 030329 and SN 2003dh makes it clear that we
must understand Type Ic SNe in order to understand GRBs.  The converse
is also true: we must understand GRBs in order to fully understand Type
Ic SNe.  It is possible that the creation of a powerful
ultra-relativistic jet as a result of the collapse of the core of a
massive star to a black hole plays a direct role in Type Ic supernova
explosions \cite{macfadyen2001}, but it is certain that the rapid
rotation of the collapsing core implied by such jets must be an
important factor in some -- perhaps many -- Type Ic supernovae.   The
result will often be a highly asymmetric explosion, whether the result
of rapid rotation alone or of the creation of powerful magnetic fields
as a result of the rapid rotation \cite{khokhlov1999}.

The large linear polarizations measured in several bright GRB
afterglows, and especially the temporal variations in the linear
polarization (see, e.g., \cite{rol2003}), provide strong evidence that
the Type Ic supernova explosions associated with GRBs are highly
asymmetric.  The recent dramatic discovery that GRB 021206 was strongly
polarized \cite{coburn2003} provides evidence that GRB jets are in fact
dominated by magnetic energy rather than hydrodynamic energy.

In addition, the X-ray afterglows of several GRBs have provided
tantalizing evidence of the presence of emission lines of
$\alpha$-particle nuclei \cite{reeves2002,butler2003}.  These emission
lines, if confirmed, provide severe constraints on models of GRBs and
Type Ic supernovae [see, e.g., \cite{lazzati2002}].  They may also
provide information on the abundances and properties of heavy elements
that have been freshly minted in the supernova explosion.

It is therefore now clear that GRBs are a unique laboratory for
studying, and are a powerful tool for understanding, Type Ic core
collapse supernovae.

Confirmation by HETE-2 of the connection between GRBs and Type Ic
supernovae has also strengthened the expectation that GRBs occur out to
redshifts $z \sim 20$, and are therefore a powerful probe of cosmology
and the early universe.

\section{Short GRBs}

Assuming that short bursts follow the star-formation rate (as long
bursts are thought to do), one can show \cite{schmidt2001} that the
peak luminosities of short bursts are essentially the same as those of
long bursts. Otherwise, little is known about the distance scale or the
nature of short GRBs.  {\it Beppo}SAX did not detect any short, hard
GRBs during its 6-year mission, despite extensive efforts.  The rapid
HETE-2 and IPN localizations of GRB 020531 \cite{lamb2003a} made
possible rapid optical ($t$ = 2-3 hours) follow-up observations.  No
optical afterglow was detected
\cite{lamb2002,miceli2002,dullighan2002}.  Chandra follow-up
observations at $t = 5$ days showed that $L_X ({\rm short})/L_x ({\rm
long}) < 0.01 -0.03$ \cite{butler2002}. These results suggest that real
time or near-real time X-ray follow-up observations of short GRBs may
be vital to unraveling the mystery of short GRBs.

\section{``Optically Dark'' GRBs}

Only $\approx$ 35\% of {\it Beppo}SAX localizations of GRBs led to the
identification of an optical afterglow.  In contrast, 13 of the 14 GRBs
localized so far by the WXM plus the SXC on HETE-2 have optical
afterglows.  HETE-2 is thus solving the mystery of ``optically dark''
bursts.

% 030115 NIR Observations
\begin{figure}[t]
\centerline{\psfig{file=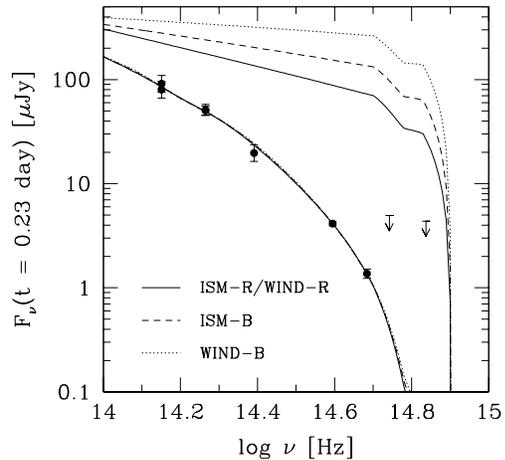,width=0.42\textwidth}} 
\caption{NIR and optical afterglow spectrum of GRB 030115, as
determined from K, H, J, i*, and r* observations.  The curve that goes
through the data points is the best-fit model, assuming extinction by
dust of the four theoretical afterglow spectra labeled "ISM-R, WIND-R,
ISM-B, WIND-B."  The amount of extinction by dust is a sensitive
function of the redshift of the burst.  The redshift of this burst has
not yet been reported; the case shown therefore assumes $z = 3.5$, the
largest redshift allowed by the observations and the one that
attributes the {\it least} amount of extinction by dust.  The amount of
extinction by dust in the optical is still substantial.  From
\cite{lamb2003b}.
\label{fig7}}
\end{figure}

Three explanations of ``optically dark'' GRBs have been widely discussed:

\begin{itemize}

\item 
The optical afterglow is extinguished by dust in the vicinity of the
GRB or in the star-forming region in which the GRB occurs (see, e.g.,
\cite{lamb2000,reichart2002}).

\item
The GRB lies at very high redshift ($z > 5$), and the optical afterglow
is absorbed by neutral hydrogen in the host galaxy and in the
intergalactic medium along the line of sight from the burst to us
\cite{lamb2000b}.

\item
Some GRBs have optical afterglows that are intrinsically very faint
(see, e.g., \cite{fynbo2001,berger2002,lcg2002}).

\end{itemize}

% 021211 followups
\begin{figure}[t]
\centerline{\psfig{file=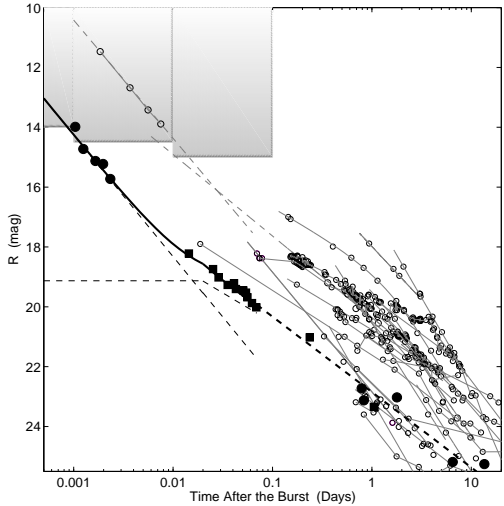,width=0.42 \textwidth}} 
\caption{Light curve of the optical afterglow of GRB 021211, compared
to those of other GRBs.  The dashed curve in the upper left-hand corner
of the figure shows the light curve of the optical afterglow of GRB
990123, while the dashed horizontal line in the left-hand middle of the
figure shows the light curve of the optical afterglow of GRB 021004. 
HETE-2 has shown that the optical afterglows of GRBs can exhibit a wide
range of behaviors in the first few hours after the burst.  From
\cite{fox2003b}.
\label{fig9}}
\end{figure}

Rapid optical follow-up observations of the HETE-2--localized burst
GRB030115 \cite{kawai2003} show that the optical afterglow of this
burst is the best case observed to date of a burst whose optical
afterglow is extinguished by dust.  Figure 3 shows the NIR and optical
afterglow spectrum of this burst and the best-fit model, assuming
extinction by dust \cite{lamb2003b}.  The amount of extinction by dust
is a sensitive function of the redshift of the burst.  Since the
redshift of this burst has not been reported as yet, the case shown in
Figure 3 assumes $z = 3.5$, the largest redshift allowed by the
observations and the one that attributes the {\it least} amount of
extinction by dust.  The amount of extinction by dust in the optical is
still substantial.

Rapid optical follow-up observations
\cite{fox2003b,park2002,li2003,wozniak2002} of the HETE-2--localized
burst GRB 021211 \cite{crew2003} show that the optical afterglow of
this burst is intrinsically much fainter than those observed
previously.  Figure 4 shows the light curve of the afterglow of GRB
021211 \cite{fox2003b}.  The transition from the reverse shock
component \cite{sari1999} to the forward shock component is clearly
visible.  Figure 4 also compares the afterglow of GRB 021211 to those
of other GRBs.  These observations show that the light curve of the
afterglow of this burst tracks those of GRBs 990123 and 030329, but is
three and six magnitudes fainter than them.

This burst would almost certainly have been classified as ``optically
dark'' were it not for its accurate, rapid localization by HETE-2. 
Upper limits or measurements of the optical afterglows of other {\it
Beppo}SAX-- and HETE-2--localized bursts suggest that they too have
afterglows that are very faint (see, e.g.,
\cite{fynbo2001,berger2002,lcg2002}).   GRBs with intrinsically faint
afterglows may therefore account for a substantial fraction of bursts
previously classified as ``optically dark.''  

The temporal behavior of the optical afterglow of the HETE-2--localized
burst GRB 021004 was nearly flat at early times -- a behavior that is
different than any seen previously and that suggests the ``central
engine'' powering the GRB continued to pour out energy long after the
burst itself was over\cite{fox2003a}.  Thus HETE-2 is making it
possible to explore the previously unknown behavior of GRB afterglows 
in the ``gap'' in time from the end of the burst to 3 - 20 hours after
the burst that existed in the {\it Beppo}SAX era.

\section{GRBs as a Probe of Cosmology and the Early Universe}

HETE-2 has decisively confirmed the connection between GRBs and the
deaths of massive stars, as we have seen.  The earliest massive stars
are thought to have formed at redshifts $z \approx 20$
\cite{gnedin1997,valagaes1999} and died soon thereafter.  Thus GRBs
may mark the moment of ``first light'' and an end to the ``dark ages''
of the universe.  Indeed, recent calculations suggest that 10-40\% of
all GRBs may lie at very high ($z > 5$) redshifts
\cite{lamb2000b,ciardi2000,bromm2002}.

% GRBs in cosmological context.
\begin{figure}[t]
\centerline{\psfig{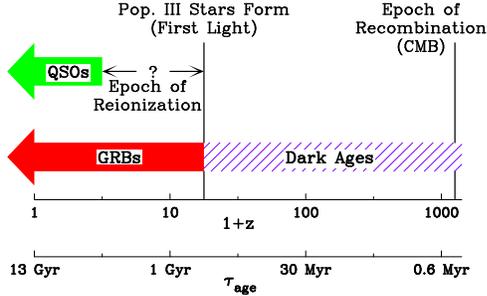}} 
\caption{Cosmological context of VHR GRBs.  Shown are the epochs of
recombination, first light, and re-ionization.  Also shown are the
ranges of redshifts corresponding to the ``dark ages''probed
by QSOs and GRBs.  From \cite{lamb2002}.
\label{fig11}}
\end{figure}

GRBs are far and away the brightest objects in the universe, with
$\gamma$-ray luminosities that are frequently 10 billion times greater
than the optical luminosities of the supernovae with which they are
associated, or of their host galaxies.  It is no surprise, then, that the
bursts are easily detectable out to redshifts $z \approx 20$ by HETE-2.

% Correlation of E_iso with z.
\begin{figure}[t]
\centerline{\psfig{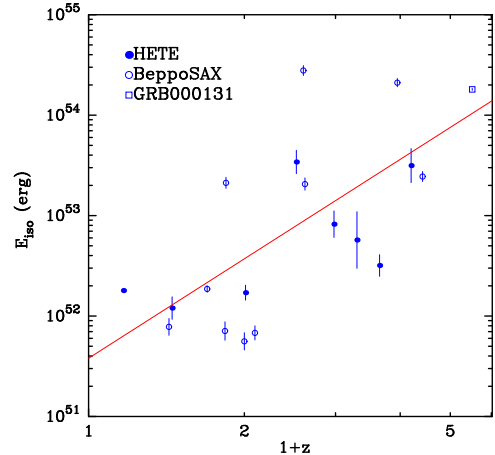}} 
\caption{Distribution of {\it Beppo}SAX and HETE-2 GRBs with known
redshifts in the ($1+z,E_{\rm iso}$)-plane.  This distribution provides
modest evidence that $E_{\rm iso}$, the isotropic-equivalent
$\gamma$-ray energy of GRBs, is correlated with redshift (the
threshold-corrected chance probability of such a correlation is 0.05). 
The HETE-2 results provide stronger evidence that $L_{\rm iso}$, the
isotropic-equivalent $\gamma$-ray luminosity of GRBs, is correlated
with redshift (the threshold-corrected chance probability of such a
correlation is $9.5 \times 10^{-3}$).  If confirmed, these results
would imply that GRBs evolve strongly with redshift.  From
\cite{lamb2003c}. \label{fig13}} \end{figure}

Somewhat surprisingly, the infrared and near-IR afterglows of GRBs are
also detectable out to very high redshifts \cite{lamb2000b}.  The
reason is that while the increase in distance and the redshift of the
spectrum tend to reduce the spectral flux in a given frequency band,
cosmological time dilation tends to increase it at a fixed time of
observation after the GRB, since afterglow intensities decrease with
time.  These effects combine to produce little or no decrease -- and
can even produce an increase -- in the spectral energy flux of GRB
afterglows beyond $z \approx 3$ \cite{lamb2000b,ciardi2000}. 
Consequently, ``optically dark,'' but ``near-infrared and infrared
bright,'' GRBs may be a powerful probe of cosmology and the early
universe.

Figure 5 places GRBs in a cosmological context \cite{lamb2002}.  At
recombination, which occurs at redshift $z = 1100$, the universe
becomes transparent. The cosmic background radiation originates at this
redshift.  Shortly afterward, the temperature of the cosmic background
radiation falls below 3000 K and the universe enters the ``dark ages''
during which there is no visible light in the universe.  ``First
light,'' which occurs at $z \approx 20$, corresponds to the epoch when
the first stars form.   Ultraviolet radiation from these first stars
and/or from the first active galactic nuclei re-ionizes the universe. 
Afterward, the universe is transparent in the ultraviolet.

Important cosmological questions that observations of GRBs and their
afterglows may address include the following \cite{lamb2000b}:
\medskip

\noindent
$\bullet$ The moment of ``first light'' and the earliest generations of
stars merely by the detection of GRBs at very high redshifts;
\medskip

\noindent
$\bullet$ The re-ionization of the universe by the shape of the red
damping wing of the Gunn-Peterson trough due to Lyman $\alpha$ in the
spectra of GRB afterglows.
\medskip

\noindent
$\bullet$ The history of metallicity growth in the universe -- in the
star-forming entities in which the bursts occur, in damped Lyman
$\alpha$ clouds, and in the Lyman $\alpha$ forest -- by observations
of metal absorption line systems in the spectra of GRB afterglows; and
\medskip

\noindent
$\bullet$ The large-scale structure of the universe at very high
redshifts ($z > 5$) by the clustering of Lyman $\alpha$ forest lines
and the metal absorption-line systems in the spectra of their
afterglows.
\medskip

The recently announced results from the Wilkinson Microwave Anisotropy
Probe indicate that re-ionization of the universe occurred at $z = 17
\pm 5$ \cite{spergel2003}, much earlier than previously thought.
Such redshifts are far beyond those that can be probed using quasars or
even Ly$\alpha$-emission galaxies.  In contrast, GRBs are expected to
occur out to the redshifts ($z \approx 20$) at which the first stars
formed, stars whose UV light evidently re-ionizes the universe.  Thus,
GRBs may provide a unique probe of the re-ionization of
the universe and the moment of ``first light.''

HETE-2 has provided evidence that the isotropic-equivalent energies
$E_{\rm iso}$ and luminosities $L_{\rm iso}$ of GRBs are correlated
with redshift \cite{lamb2003} (see Figure 6).  The apparent 
correlations imply that GRBs evolve strongly with redshift -- bursts
at $z = 5$ appear to be $\sim$ 400 times more luminous than those at $z = 0$. 
These results strengthen evidence to this effect from analyses of
the BATSE catalog of GRBs, using the variability of burst time
histories as an estimator of burst luminosities (and therefore
redshifts)
\cite{fenimore2000,reichart2001a,reichart2001b,lloyd-ronning2002}, and
from an analysis of {\it Beppo}SAX bursts only \cite{amati2002}.

Strong evolution of $E_{\rm iso}$ and $L_{\rm iso}$ with $z$ would have
profound implications for the nature of GRB progenitors and for the
use of GRBs a a probe of cosmology and the early universe.  The HETE-2
results suggest that GRBs at redshifts $z$ = 10 - 20 may be much more
luminous -- and therefore easier to detect -- than has been thought,
making GRBs a more powerful probe of the very high redshift universe.

% Hardness histogram for HETE-2 GRBs.
\begin{figure}[t]
\centerline{\psfig{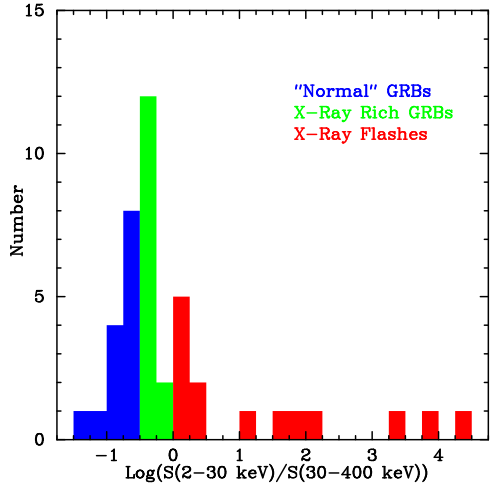}} 
\caption{Hardness histogram for HETE-2 GRBs.  Shown are GRBs
(blue histogram), X-ray-rich GRBs (green histogram), and XRFs (red
histogram).  From \cite{sakamoto2003b}.
\label{fig14}}
\end{figure}

\section{Nature of X-Ray Flashes and X-Ray-Rich GRBs}

Two-thirds of all HETE-2--localized bursts are either ``X-ray-rich'' or
XRFs, and one-third are XRFs (see Figure 7).\footnote{We define
``X-ray-rich'' GRBs and XRFs as those events for which $\log
[S_X(2-30~{\rm kev})/S_\gamma(30-400~{\rm kev})] > -0.5$ and 0.0,
respectively.}  These events have received increasing attention in the
past several years \cite{heise2000,kippen2002}, but their nature
remains largely unknown.

% Fluence versus E_pk, by hardness.
\begin{figure}[t]
\centerline{\psfig{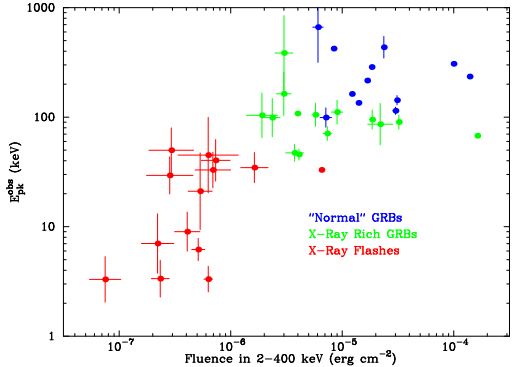}} 
\caption{Distribution of HETE-2 bursts in the [$S(2-400 {\rm keV}),
E^{\rm obs}_{\rm peak}$]-plane, showing XRFs (red), X-ray-rich GRBs
(green), and GRBs (blue).    From \cite{sakamoto2003b}.
\label{fig16}}
\end{figure}

% Amati et al. (2002) relation.
\begin{figure}[t]
\centerline{\psfig{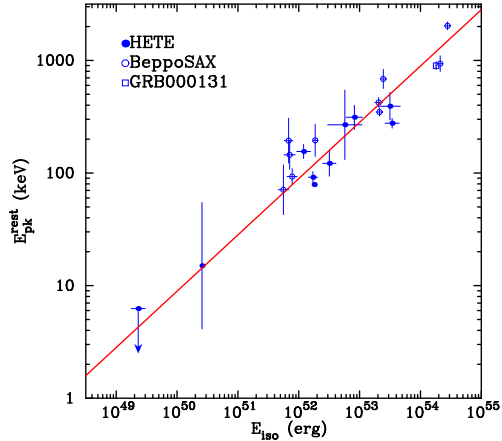}} 
\caption{Distribution of HETE-2 and BeppoSAX bursts in the ($E_{\rm
iso}$,$E_{\rm peak}$)-plane, where $E_{\rm iso}$ and $E_{\rm peak}$ are
the isotropic-equivalent GRB energy and the peak of the GRB spectrum in
the source frame.  The HETE bursts confirm the relation between $E_{\rm
iso}$ and $E_{\rm peak}$ found by Amati et al. (2002), and extend it by
a factor $\sim 300$ in $E_{\rm iso}$.  The bursts with th lowest and
second-lowest values of $E_{\rm iso}$ are XRFs 020903 and 030723.
From \cite{lamb2003c}.
\label{fig17}}
\end{figure}

Clarifying the nature of XRFs and X-ray-rich GRBs, and their connection
to GRBs, could provide a breakthrough in our understanding of the
prompt emission of GRBs.  The spectrum of the HETE-2--localized event
XRF 020903 \cite{sakamoto2003a} gave an upper limit $E^{\rm obs}_{\rm
peak} <$ 5 keV (99.7\% confidence level) making this event one of the
softest bursts seen so far by HETE-2.  Follow-up observations made
possible by the HETE-2 localization identified the likely optical
afterglow of the XRF \cite{soderberg2002}. Later observations
determined that the optical transient occurred in a star-forming galaxy
at a distance $z = 0.25$ \cite{soderberg2002,chornock2002}; both of
these properties are typical of GRB host galaxies.

Analyzing 42 X-ray-rich GRBs and XRFs seen by FREGATE and/or the WXM
instruments on HETE-2, \cite{sakamoto2003b} find that the XRFs, the
X-ray-rich GRBs, and GRBs form a continuum in the [$S_\gamma(2-400~{\rm
kev}), E^{\rm obs}_{\rm peak}$]-plane (see Figure 8).  This result
strongly suggests that all three kinds of events are the same
phenomenon.

Furthermore, \cite{lamb2003c} have placed 9 HETE-2 GRBs with known
redshifts and 2 XRFs with known redshifts or strong redshift
constraints in the ($E_{\rm iso}, E_{\rm peak}$)-plane (see Figure 9). 
Here $E_{\rm iso}$ is the isotropic-equivalent burst energy and $E_{\rm
peak}$ is the energy of the peak of the burst spectrum, measured in the
source frame.  The HETE-2 bursts confirm the relation between $E_{\rm
iso}$ and $E_{\rm peak}$ found by Amati et al.  \cite{amati2002} for
GRBs and extend it down in $E_{\rm iso}$ by a factor of 300.  The fact
that XRF 020903, one of the softest events localized by HETE-2 to date,
and XRF 030723, the most recent XRF localized by HETE-2, lie squarely
on this relation  \cite{sakamoto2003a,lamb2003c} provides additional
evidence that XRFs and GRBs are the same phenomenon.  However,
additional redshift determinations are clearly needed for XRFs with 1
keV $< E_{\rm peak} < 30$ keV energy in order to confirm these results.

\section{Conclusions}

The HETE-2 mission has been highly productive.  It has observed more
than 250 GRBs so far.  It is currently localizing 25 - 30 GRBs per
year, and has localized 43 GRBs to date.  Twenty-one of these
localizations have led to the detection of X-ray, optical, or radio
afterglows, and as of now, 11 of the bursts with afterglows have
redshift determinations.  HETE-2 has also observed more than 45 bursts
from soft gamma-ray repeaters, and more than 700 X-ray bursts.

HETE-2 has confirmed the connection between GRBs and Type Ic
supernovae, a singular achievement and certainly one of the scientific
highlights of the mission so far.  It has provided evidence that the
isotropic-equivalent energies and luminosities of GRBs are 
correlated with redshift, implying that GRBs and their progenitors
evolve strongly with redshift.  Both of these results have profound
implications for the nature of GRB progenitors and for the use of GRBs
as a probe of cosmology and the early universe.

HETE-2 has placed severe constraints on any X-ray or optical afterglow
of a short GRB.  It has made it possible to explore the previously
unknown behavior optical afterglows at very early times, and has opened
up the era of high-resolution spectroscopy of GRB optical afterglows.
It is also solving the mystery of ``optically dark'' GRBs, and
revealing the nature of X-ray flashes.

\end{document}